\documentclass[12pt]{article}
\usepackage[english,german,french,polish]{babel}
\usepackage[T1]{fontenc}
\usepackage{amsfonts}

\begin{document}

\selectlanguage{english}

\baselineskip 0.76cm
\topmargin -0.6in
\oddsidemargin -0.1in

\let\ni=\noindent

\renewcommand{\thefootnote}{\fnsymbol{footnote}}

\pagestyle {plain}

\setcounter{page}{1}

\pagestyle{empty}

~~~

\begin{flushright}
IFT--05/22
\end{flushright}

\vspace{0.5cm}

{\large\centerline{\bf Can the nondiagonal charged-lepton mass matrix}}

{\large\centerline{\bf improve the tripartite neutrino mixing?}}

\vspace{0.5cm}

{\centerline {\sc Wojciech Kr\'{o}likowski}}

\vspace{0.3cm}

{\centerline {\it Institute of Theoretical Physics, Warsaw University}}

{\centerline {\it Ho\.{z}a 69,~~PL--00--681 Warszawa, ~Poland}}

\vspace{0.6cm}

{\centerline{\bf Abstract}}

\vspace{0.2cm}

It is shown that the off-diagonal corrections in a realistic charged-lepton mass matrix, introduced 
previously by the author, cannot improve much the tripartite neutrino mixing. In fact, $s^2_{12} = 
1/3 - 0.005 = 0.328 \simeq 1/3 $ for $s^2_{12}$ improved in this way, while $s^2_{12\,{\rm exp}}
= 0.314 = 1/3 - 0.019$ for the central value of experimental $s^2_{12\,{\rm exp}}$.

\vspace{0.2cm}

\ni PACS numbers: 14.60.Pq , 12.15.Ff  .

\vspace{0.6cm}

\ni September 2005 

\vfill\eject

~~~~

\pagestyle {plain}

\setcounter{page}{1}

The neutrino mixing matrix $ U = (U_{\alpha i})\;(\alpha = e, \mu, \tau$ and $i = 1,2,3)$, connecting the flavor neutrinos $\nu_e , \nu_\mu , \nu_\tau $ and mass neutrinos $\nu_1 , \nu_2 , \nu_3 $ through the unitary transformation

\begin{equation}
\nu_\alpha  = \sum_i U_{\alpha i}\, \nu_i \;,
\end{equation}

\ni is experimentally consistent with the familiar bilarge form

\begin{equation}
U = \left( \begin{array}{rrr} c_{12}\, & s_{12}\, & 0\, \\ - \frac{1}{\sqrt2} s_{12} & \frac{1}{\sqrt2} c_{12} & \frac{1}{\sqrt2} \\ \frac{1}{\sqrt2} s_{12} & -\frac{1}{\sqrt2} c_{12} & \frac{1}{\sqrt2}  \end{array} \right) \;,
\end{equation}

\vspace{0.2cm}

\ni where (at $\pm 2\sigma$) $s^2_{12} = 0.314 \left( 1^{+0.18}_{-0.15}\right)$ and $s^2_{23} = 0.44 \left(1^{+0.41}_{-0.22}\right)$ [1], the latter being put 1/2, while $U_{e3} = s_{13}\exp(-i \delta)$ is neglected due to the nonobservation of $\bar{\nu_e} \rightarrow \bar{\nu_e}$ oscillations in the Chooz reactor experiment [2] ($s^2_{13} = 0.9^{+2.3}_{-0.9}\times 10^{-2}$ [1] is put zero).

If the so-called tripartite neutrino mixing [3] was true, then Eq. (2) would take the form

\begin{equation}
U = \left( \begin{array}{rrr} \sqrt{\frac{2}{3}} & \frac{1}{\sqrt3} & 0 \\ -\frac{1}{\sqrt6} & \frac{1}{\sqrt3} & \frac{1}{\sqrt2} \\ \frac{1}{\sqrt6} & -\frac{1}{\sqrt3} & \frac{1}{\sqrt2} \end{array} \right) \;.
\end{equation}

\vspace{0.2cm}

\ni In this case $s^2_{12} = 1/3 = 0.333 $.

Mainly due to E. Ma [4], this recently popular neutrino mixing is often associated with the 
non-Abelian discrete group $A_4$, the group of even permutations of four objects (also known as the symmetry group of a regular tetrahedron). At any rate, the effective neutrino mass matrix $M^{(\nu)} = \left( M^{(\nu)}_{\alpha \beta}\right)$ can be fully expressed as a linear combination of $3\times 3$ matrices generating the $A_4$ group, and this specific combination describes the way of breaking the $A_4$ symmetry by three active neutrinos ({\it cf.} the last Ref. [4]). Twelve $3\times 3$ generators of the $A_4$ group are connected through three linear identities, and can be linearly expressed by nine Gell-Mann matrices ${\bf 1}, \lambda_1,\ldots, \lambda_8$ generating a formal, horizontal $U(3)$ group.

In this note, we assume tentatively that the neutrino diagonalizing matrix $ U^{(\nu)} = (U^{(\nu)}_{\alpha i})$, {\it i.e.} such that $U^{(\nu)\,\dagger} M^{(\nu)} U^{(\nu)} = $ diag$(m_1, m_2, m_3)$, has the tripartite form (3):
 
\begin{equation}
U^{(\nu)}  = \left( \begin{array}{rrr} \sqrt\frac{2}{3} & \frac{1}{\sqrt3} & 0 \\ -\frac{1}{\sqrt6} & \frac{1}{\sqrt3} & \frac{1}{\sqrt2} \\ \frac{1}{\sqrt6} & -\frac{1}{\sqrt3} & \frac{1}{\sqrt2} \end{array} \right) \;.
\end{equation}

\vspace{0.3cm}

\ni The neutrino mixing matrix $U$ gets in general the structure

\begin{equation}
U = U^{(e)\,\dagger} U^{(\nu)} \,,
\end{equation}

\ni where $U^{(e)} = \left(U^{(e)}_{\alpha \beta} \right)$ is the charged-lepton diagonalizing matrix. For the latter we assume the form described in Ref. [5]:

\begin{equation}
U^{(e)} = \left(\begin{array}{ccc} 1 & \frac{2}{29} \frac{\alpha}{\stackrel{\circ}{m}_\mu - \stackrel{\circ}{m}_e} & 0 \\ -\frac{2}{29} \frac{\alpha}{\stackrel{\circ}{m}_\mu - \stackrel{\circ}{m}_e} & 1 & \frac{8\sqrt3}{29} \frac{\alpha}{\stackrel{\circ}{m}_\tau - \stackrel{\circ}{m}_\mu} \\ 0 & -\frac{8\sqrt3}{29} \frac{\alpha}{\stackrel{\circ}{m}_\tau - \stackrel{\circ}{m}_\mu} & 1 \end{array}\right) + O(\frac{\alpha^2}{\mu^2})\,,
\end{equation}

\vspace{0.3cm}

\ni where $\alpha > 0$ is a constant (massdimensional). The form (6) is valid in the first perturbative order in $\alpha $, if the charged-lepton mass matrix $M^{(e)} = \left(M^{(e)}_{\alpha \beta}\right)$ is conjectured to be parametrized in the following way :

\begin{equation}
M^{(e)} = \left(\begin{array}{ccc} \stackrel{\circ}{m}_e & \frac{2\alpha}{29} & 0 \\ \frac{2\alpha}{29} & \stackrel{\circ}{m}_\mu & \frac{8\sqrt{3}\alpha}{29} \\ 0 & \frac{8\sqrt{3}\alpha}{29} & \stackrel{\circ}{m}_\tau \end{array}\right) \,,
\end{equation}

\vspace{0.2cm}

\ni where

\begin{equation}
\stackrel{\circ}{m}_e = \frac{\mu}{29}\varepsilon \;\;,\;\; \stackrel{\circ}{m}_\mu = \frac{\mu}{29} \frac{4}{9} (80 + \varepsilon) \;\;,\;\; \stackrel{\circ}{m}_\tau  = \frac{\mu}{29} \frac{24}{25} (624 + \varepsilon) \,,
\end{equation}

\vspace{0.2cm}

\ni while $\mu > 0$ and $\varepsilon > 0$ are two constants (massdimensional and dimensionless, respectively). A motivation for the form (7) of the charged-lepton mass matrix the interested reader may find in Appendix in Ref. [5].

Performing the diagonalization $U^{(e) \dagger} M^{(e)}U^{(e)} =$ diag$(m_e, m_\mu, m_\tau)$ of the charged-lepton mass matrix $M^{(e)}$, we obtain the masses $\,m_\alpha = \stackrel{\circ}{m}_\alpha\, + \,\delta m_\alpha\;\,(\alpha = e, \mu, \tau)\,$, where $ \stackrel{\circ}{m}_\alpha $ are given as in Eqs. (8) and $\,\delta m_\alpha\,$ are calculated to be [5] (here, $\,\delta m_e + \delta m_\mu + \delta m_\tau = 0\,$):

\vfill\eject

\begin{eqnarray}
\delta m_e & = & -\frac{36}{29}\left(\frac{\alpha}{\mu}\right)^2 \frac{\mu}{320 - 5 \varepsilon} \,, \nonumber \\
\delta m_\mu & = & \frac{36}{29}\left(\frac{\alpha}{\mu}\right)^2 \frac{\mu}{320 - 5 \varepsilon} - \frac{10800}{29}\left(\frac{\alpha}{\mu}\right)^2 \frac{\mu}{31696 + 29 \varepsilon} \,, \nonumber \\
\delta m_\tau & = & \frac{10800}{29}\left(\frac{\alpha}{\mu}\right)^2 \frac{\mu}{31696 + 29 \varepsilon}
\end{eqnarray}

\ni in the second perturbative order in $\alpha$. Then, $U^{(e)}$ is also obtained in the second perturbative order in $\alpha$, giving the form (6) in the first order.

Making use of Eqs. (8) and (9), we calculate:

\begin{equation}
m_\tau = \frac{6}{125}\left[351\left(m_\mu - \delta m_\mu\right) - 136\left(m_e - \delta m_e\right) \right] +  \delta m_\tau = \left( 1776.80 + 10.2112 \frac{\alpha^2 }{\mu^2 }\,\right)\;{\rm MeV}  
\end{equation}

\ni and 

\begin{eqnarray}
\mu & = & \frac{29}{320}\left[ 9\left(m_\mu - \delta m_\mu\right) - 4\left(m_e - \delta m_e\right) \right] =  \left[85.9924  + O\left( \frac{\alpha^2 }{\mu^2 } \right)\,\right] \;{\rm MeV}   \;, \nonumber \\ 
\varepsilon & = & \frac{320 (m_e - \delta m_e)}{9(m_\mu - \delta m_\mu) - 4 (m_e - \delta m_e)} = 0.172329 + O\left( \frac{ \alpha^2}{ \mu^2}\right) \;,
\end{eqnarray}

\ni where the experimental values of $m_e = 0.5109989$ MeV and $m_\mu = 105.65837$ MeV [6] are inserted as an input. In Eqs. (10) and (11), the nonperturbed parts are: 

\begin{equation}
\stackrel{\circ}{m}_\tau = \frac{6}{125}\left(351m_\mu  - 136 m_e \right) = 1776.80 \;{\rm MeV}   
\end{equation}

\ni and 

\begin{eqnarray}
\stackrel{\circ}{\mu} & = & \frac{29}{320}\left( 9m_\mu - 4 m_e \right) =  85.9924 \;{\rm MeV} \;, \nonumber \\ 
\stackrel{\circ}{\varepsilon} & = & \frac{320 m_e}{9m_\mu - 4 m_e} = 0.172329 \;.
\end{eqnarray}

We can see from Eq. (10) that even for $\alpha = 0$ the predicted value of $m_\tau $ agrees very well with its experimental value $m_\tau = 1776.99^{+0.29}_{-0.26}$ MeV [6]. To estimate $\alpha^2/\mu^2 $, we can use this experimental value as an additional input, obtaining then the estimation

\begin{equation}
\frac{\alpha^2}{\mu^2} = 0.019^{+0.028}_{-0.019} 
\end{equation}

\ni that, consistently, turns out to be small (the value $\alpha = 0$ is not experimentally excluded). Thus, the form (7) of the charged-lepton mass matrix may be considered as realistic.

Though the parametrization (7) has essentially the empirical character, there is a speculative background for it [5, 7] based on a K\"{a}hler-like extension of the Dirac equation ({\it i.e.} on the extended Dirac's square-root procedure), and, on the other hand, on the Pauli principle realized in an intrinsic way for additional bispinor indices implied by such an extended Dirac equation (and treated as physical objects: spin-1/2 algebraic partons). This Pauli principle restricts the number of lepton and quark genearations exactly to three.

Now, making use of Eqs. (5) and (2), (4), (6), we obtain in the first perturbative order in $\alpha$ the following prediction:

\begin{eqnarray}
U_{e\,1} & = & c_{12} = \sqrt{\frac{2}{3}}\left( 1 + \frac{1}{29}\frac{\alpha}{m_\mu - m_e} \right) = 0.820 \,,\nonumber \\
U_{e\,2} & = & s_{12} = \frac{1}{\sqrt3}\left(1 -\frac{2}{29} \frac{\alpha}{m_\mu - m_e} \right) = 0.573
\;,   
\end{eqnarray}

\ni where $\alpha/(m_\mu -m_e) = 0.113$ with the use of the central value of $\alpha^2/\mu^2 $ in Eq. (14) and of the value $\stackrel{\circ}{\mu}$ as in Eq. (13) for $\mu $. Hence,

\begin{equation}
c^2_{12} = 0.672 \;,\; s^2_{12} = 0.328 
\end{equation}

\ni and so, the deviation of this prediction from the tripartite neutrino mixing is small:

\begin{equation}
c^2_{12} - \frac{2}{3} = 0.672 - 0.667 = 0.005 \simeq 0 \;,\; s^2_{12} - \frac{1}{3} = 0.328 - 0.333 = -0.005 \simeq 0 \,,
\end{equation}

\ni while its deviation from the experiment is larger:

\begin{equation}
s^2_{12} - s^2_{12\,{\rm exp}} = 0.328 - 0.314 = 0.014
\end{equation}

\ni for the central value of experimental $s^2_{12\,{\rm exp}}$. The deviation of the tripartite neutrino mixing from the experiment is still larger:

\begin{equation}
\frac{1}{3} - s^2_{12\,{\rm exp}} = 0.333 - 0.314 = 0.019
\end{equation}

\ni for this central value.

Thus, we can conclude that the off-diagonal corrections in a realistic charged-lepton mass matrix (7) introduced previously [5] are negligible as far as neutrino oscillations are concerned: $U = 
U^{(e)\,\dagger} U^{(\nu)} \simeq U^{(\nu)}$, though they play a correcting role in the precise prediction of $m_\tau $ from the input of experimental $m_e $ and $m_\mu$. Thus, the tentative tripartite form (4) of $U^{(\nu)}$ may differ considerably from reality, since such a tripartite $U^{(\nu)}$ may be not close enough to the real $U^{(\nu)} \simeq U^{(e)\,\dagger} U^{(\nu)} = U$ given in Eq. (2) with the use of experimental values of $c_{12}$ and $s_{12}$.

Recently, deviations from the tripartite (= tribimaximal) neutrino mixing were systematically discussed in Ref. [8]. Also Ref. [9] deals with this problem.

\vfill\eject

~~~~

\vspace{0.4cm}

{\centerline{\bf References}}

\vspace{0.4cm}

{\everypar={\hangindent=0.7truecm}
\parindent=0pt\frenchspacing

{\everypar={\hangindent=0.7truecm}
\parindent=0pt\frenchspacing

[1]~G.L. Fogli, E. Lisi, A. Marrone and A. Palazzo, {\tt hep--ph/0506083}.

\vspace{0.15cm}

[2]~M.A. Apollonio {\it et al.} (Chooz Collaboration), {\it Eur. Phys. J.} {\bf C 27}, 331 (2003).

\vspace{1.5mm}

[3]~P.F. Harrison, D.H. Perkins and W.G.~Scott, {\it Phys. Lett.} {\bf B 458}, 79 (1999); {\it Phys. Lett.} {\bf B 530}, 167 (2002); Z.-Z.~Xing. {\it Phys. Lett.} {\bf B 553}, 85 (2003); X.-G.~He and A.~Zee, {\it Phys. Lett.} {\bf B 560}, 87 (2003);  {\it Phys. Rev.} {\bf D 68}, 037302 (2003); E.Ma, {\it Phys. Lett.} {\bf B 583}, 157 (2004).

\vspace{1.5mm}

[4]~E. Ma and G. Rajasekaran, {\it Phys. Rev.} {\bf D 64}, 113012 (2001), and references therein; K.S.~Babu, E.~Ma and J.W.~Valle, {\it Phys. Lett.} {\bf B 552}, 207 (2003); E.~Ma, {\it Phys. Rev.} {\bf D 70}, 031901(R) (2004); {\tt hep-ph/0505209}; K.S.~Babu and X.-G.~He, {\tt hep-ph/0507217}; G.~Altarelli, {\tt hep-ph/0508053}; A.~Zee, {\tt hep-ph/0508278}; {\it cf.} also W.~Kr\'{o}likowski {\tt hep-ph/0501008}.

\vspace{0.15cm}

[5]~W. Kr\'{o}likowski, {\it Acta Phys. Pol.} {\bf B 32}, 2961 (2001) [{\tt hep-ph/0108157}], and earlier references therein.

\vspace{0.15cm}

[6]~Particle Data Group, {\it ~Review of Particle Physics}, {\bf B 592}, 1 (2004).

\vspace{0.15cm}

[7]~For a recent presentation {\it cf.} W. Kr\'{o}likowski, {\it Acta Phys. Pol.} {\bf B 33}, 2559 (2002) [{\tt hep-ph/0203104}], and earlier references therein; {\it cf.} also W. Kr\'{o}likowski, {\it Acta Phys. Pol.} {\bf B 36}, 2051 (2005) [{\tt hep-ph/0503074}] and {\tt hep-ph/0504256}.

\vspace{0.15cm}

[8]~F. Plantinger and W. Rodejohann, {\tt hep-ph/0507143}.

\vspace{0.15cm}

[9]~S. Antusch and S.F. King, {\tt hep--ph/0508044}.

\vfill\eject

\end{document}